\documentclass[twocolumn,showpacs,preprintnumbers,amsmath,amssymb,superscriptaddress,pra]{revtex4-1}

\usepackage{graphicx}
\usepackage{dcolumn}
\usepackage{bm}
\usepackage{graphics}
\usepackage{subfigure}
\usepackage{graphicx}
\usepackage{epsfig}
\usepackage{epstopdf}
\usepackage{color}
\usepackage{soul}

\usepackage[dvipsnames]{xcolor}

\begin{document}

\title{Parity effect in a mesoscopic Fermi gas}

\author{Johannes Hofmann}
\affiliation{Condensed Matter Theory Center and Joint Quantum Institute, Department of Physics, University of Maryland, College Park, Maryland 20742-4111 USA}

\author{Alejandro M. Lobos}
\affiliation{Condensed Matter Theory Center and Joint Quantum Institute, Department of Physics, University of Maryland, College Park, Maryland 20742-4111 USA}
\affiliation{Instituto de F{\'i}sica Rosario, CCT-CONICET, Blvd. 27 de Febrero 210 bis, Rosario, Santa Fe C.P. 2000, Argentina}

\author{Victor Galitski}
\affiliation{Condensed Matter Theory Center and Joint Quantum Institute, Department of Physics, University of Maryland, College Park, Maryland 20742-4111 USA}
\affiliation{School of Physics, Monash University, Melbourne, Victoria 3800, Australia}

\date{\today}

\begin{abstract}
We develop a quantitative analytic theory that accurately describes the odd-even effect observed experimentally in a one-dimensional, trapped Fermi gas with a small number of particles [G. Z\"urn \emph{et al.}, Phys. Rev. Lett. {\bf 111}, 175302 (2013)]. We find that  the underlying physics is similar to the parity effect known to exist in ultrasmall mesoscopic superconducting grains and atomic nuclei. However, in contrast to superconducting nanograins, the density (Hartree) correction dominates over the superconducting pairing fluctuations and leads to a much more pronounced odd-even effect in the mesoscopic, trapped Fermi gas. We calculate the corresponding parity parameter and separation energy using both perturbation theory and a path integral framework in the mesoscopic limit, generalized to account for the effects of the trap, pairing fluctuations, and Hartree corrections. Our results are in an excellent quantitative agreement with experimental data and exact diagonalization. Finally, we discuss a few-to-many particle crossover between the perturbative mesoscopic regime and non-perturbative many-body physics that the system approaches in the thermodynamic limit.
\end{abstract}

\maketitle

Our understanding of quantum systems is usually firmly rooted in either a few-body picture, where exact solutions of a few-particle Schr\"odinger equation exist, or a many-body picture, where the system can be described in a (quantum) statistical framework. In between these two limits lies the mesoscopic regime, where finite particle number and confinement have a strong effect on the system's properties. Mesoscopic systems occur naturally, for example, in nuclear physics, where a finite number of protons and neutrons form an atomic nucleus, or they can be engineered, such as in semiconducting quantum dots~\cite{alivisatos96,reimann02} or superconducting nanograins~\cite{Lagarge93,ralph95,black96,ralph97}. For attractively interacting mesoscopic Fermi systems, a key effect is that the ground-state energy is not a strictly convex function of the particle number, but the interaction can cause some configurations to have lower binding energy (and thus enhanced stability) relative to others~\cite{matveev97,lee07}. For example, this implies an enhanced stability of nuclei with a ``magic number'' of constituents. A related effect exists for superconducting nanograins~\cite{matveev97,vondelft01a,vondelft01b}: the binding energy of systems with an even number of spin-up and spin-down fermions (even-number parity) is enhanced compared to odd particle number systems with an unpaired fermion (odd-number parity). This parity effect is a hallmark of mesoscopic superconductor systems and can be quantified by the so-called parity or ``Matveev-Larkin'' parameter~\cite{matveev97,vondelft01a,vondelft01b}, which denotes the excess energy of an odd parity state relative to the mean of the neighboring fully paired even parity states:
\begin{equation}
\Delta_P =E_{2l+1} - \frac{1}{2} \left(E_{2l} + E_{2l+2}\right) , \label{eq:parityparameter}
\end{equation}
where $E_{2l+1}$ denotes the ground-state energy of a fermion system with odd total particle number $2l+1$. For noninteracting systems, the parity parameter~\eqref{eq:parityparameter} vanishes, and it is positive if there is a parity effect. 

In this Rapid Communication, we study mesoscopic one-dimensional Fermi quantum gases, and establish a rigorous connection with well-known mesoscopic superconducting systems. Our work is motivated by recent progress in quantum gas experiments which can deterministically prepare systems of few fermions in a harmonic one-dimensional trap~\cite{serwane11}. These systems were studied for repulsive~\cite{zuern12} and attractive~\cite{zuern13} interactions and spin-balanced~\cite{zuern13} and spin-imbalanced configurations~\cite{wenz13}, and used to simulate models of quantum magnetism~\cite{volosniev14,murmann15a,murmann15b}. Motivated by a recent experiment~\cite{zuern13}, here we study a spin-balanced few-fermion system with attractive interaction in a harmonic trap, i.e., ensembles which contain an equal number of spin-up and spin-down fermions for a total even particle number, and a single unpaired fermion for an odd total particle number. In Ref.~\cite{zuern13}, following the preparation of an ensemble with a definite particle number, the trapping potential was tilted for a variable time, allowing fermions to tunnel out of the trap. From the  tunneling times obtained in the experiment, a separation energy was extracted~\cite{rontani12,zuern12thesis,zuern13,rontani13}, which is defined as
\begin{equation}
E_{\rm sep}(N) = (E_N - E_N^0) - (E_{N-1} - E_{N-1}^0) , \label{eq:Esep}
\end{equation}
where $E_N$ ($E_N^0$) is the ground-state energy of the interacting (noninteracting) system with $N$ particles. At zero temperature, $E_N^0$ is obtained by filling the lowest harmonic oscillator levels up to the Fermi level: for even total particle number $2l$, the states $j=0,\ldots,l-1$ are occupied by pairs of spin-up and spin-down fermions. For odd total particle number $2l+1$, the level $l$ contains an additional unpaired fermion. The parity effect is manifested in the separation energy in the form of an odd-even oscillation, where the separation energy of an odd particle number state is smaller than the separation energy of an even particle number state. The experiment~\cite{zuern13} has been analyzed theoretically using exact diagonalization for small particle number~\cite{damico15,sowinski15,pecak15}. However, for larger numbers of particles, exact diagonalization is beyond computational reach and different theoretical approaches are necessary. Recent numerical works compute ground-state properties using Monte Carlo methods for even fermion numbers up to $N=20$~\cite{berger15} and coupled-cluster methods for up to $N=80$~\cite{grining15,grining15b}. In this paper, we employ analytical methods, which allow a direct physical interpretations of the experimental results and provide complementary information to numerical works. Pairing in higher dimensions has been considered in~\cite{bruun02,heiselberg02,heiselberg03,viverit04}.
 
In the following, we analyze the mesoscopic pairing problem, focusing on the weak-interaction limit which corresponds to the experimental situation~\cite{zuern13}. The parity parameter takes a fundamentally distinct form in the few-body and the many-body limits, interpolating between a simple perturbative form and a manifestly nonperturbative many-body expression. We estimate a critical particle number which marks the crossover between the mesoscopic and the macroscopic regime, finding that this quantity scales exponentially with the interaction strength, which suggests that the mesoscopic description persists over a wide range of particle number. Our theory is in accurate quantitative agreement with the experiment~\cite{zuern13} and provides a theoretical framework to study the mesoscopic regime where $N\gg 1$, which is of fundamental interest to understand the emergence of superfluidity and superconductivity in physical systems.

The Hamiltonian of a two-component Fermi gas in one dimension is (we set $\hbar=1$)
\begin{equation}
H = \int dx \, \biggl[\sum_\sigma \psi^\dagger_\sigma \Bigl(-\frac{\partial_x^2}{2m} + V(x)\Bigr) \psi_\sigma - g_1 \psi_\uparrow^\dagger \psi_\downarrow^\dagger \psi_\downarrow^{} \psi_\uparrow^{} \biggr] . \label{eq:fullH}
\end{equation}
Here, $\psi_\sigma(x)$ annihilates a fermion at $x$ with mass $m$ and spin $\sigma$, $V(x) = m\omega^2x^2/2$ is the harmonic trapping potential with frequency $\omega$, and $g_1>0$ is related to the effective attractive scattering length $a_1$ via $g_1= 2/ma_1$. We write the continuum model~\eqref{eq:fullH} in an oscillator basis by expanding the fermion operators in terms of simple harmonic oscillator states $\psi_\sigma(x) = \sum_{j=0}^\infty c_{j\sigma} \phi_j(x)$, where $\phi_j(x)$ is a normalized harmonic oscillator wavefunction with energy $\varepsilon_j = \omega (j+1/2)$ and the operator $c_{j\sigma}$ annihilates a fermion with spin $\sigma$ in state $j$. The Hamiltonian in oscillator space is
\begin{equation}
H = \sum_{j\sigma} \varepsilon_j^{} c_{j\sigma}^\dagger c_{j\sigma}^{} - 
g_1 l_{\rm ho}^{-1} \sum_{ijkl} w_{ijkl} c_{i\uparrow}^\dagger c_{j\downarrow}^\dagger c_{k\downarrow}^{} c_{l\uparrow}^{} ,\label{eq:Hosc}
\end{equation}
where $l_{\rm ho} = \sqrt{1/m\omega}$ denotes the harmonic oscillator length. The coupling is now state-dependent with an effective interaction strength set by the overlap integral $w_{ijkl} = l_{\rm ho} \int dx \, \phi_i \phi_j \phi_k \phi_l$.

The theory in Eq.~\eqref{eq:fullH} can be solved in the absence of a trapping potential~\cite{fuchs04,feiguin12,Guan13}. In the thermodynamic limit of a large system size $L$ and large particle number $N$ with constant density $n = N/L$, the parity parameter corresponds to half the spin gap, which for small interaction strength is $\Delta_P = \tfrac{8}{\pi} \varepsilon_F \sqrt{\gamma_{\rm hom}/\pi} e^{- \pi^2/2\gamma_{\rm hom}}$~\cite{fuchs04}, where $\gamma_{\rm hom} = m g_1/n \ll 1$ is the interaction strength of the homogeneous system. For the trapped system, we expect that in the macroscopic limit of large particle number, the parity parameter is (in the Thomas-Fermi approximation) given by its minimum value at the trap center where the local density is $n_0 = 2 \sqrt{N}/\pi l_{\rm ho}$. This gives a parity parameter~\cite{grining15}
\begin{equation}
\tilde{\Delta} = \Delta_P(N\to \infty) = \frac{4 N \omega}{\pi} \sqrt{\frac{\gamma}{\pi}} e^{- \pi^2/2\gamma} , \label{eq:macroparity}
\end{equation}
where the dimensionless interaction strength is
\begin{equation}
\gamma = \frac{\pi g_1}{2 \sqrt{N} \omega l_{\rm ho}} .
\end{equation}
Equation~\eqref{eq:macroparity} is a manifestly nonperturbative expression. Note that despite the exponential suppression with $\gamma$, the parity parameter $\Delta_P$ scales with the Fermi energy. The macroscopic limit is therefore characterized by $\tilde{\Delta} \gg \omega$. By contrast, in the mesoscopic limit of small particle number where $\tilde{\Delta} \ll \omega$, we expect simple perturbation theory to hold. This is reminiscent of the Anderson criterion that marks the vanishing of superconductivity if the level spacing of a grain is larger than the bulk superconducting gap~\cite{anderson59}. Clearly, the crossover from a few to many particles is manifest in the parity parameter. The expression~\eqref{eq:macroparity} is extensive with particle number for constant $\gamma$, indicating that the crossover should be studied while keeping $\gamma$ fixed, i.e., imposing $g_1 \sim \sqrt{N}$. In the following, we consider the regime where $\gamma\ll 1$.

\begin{figure}[t]
\includegraphics{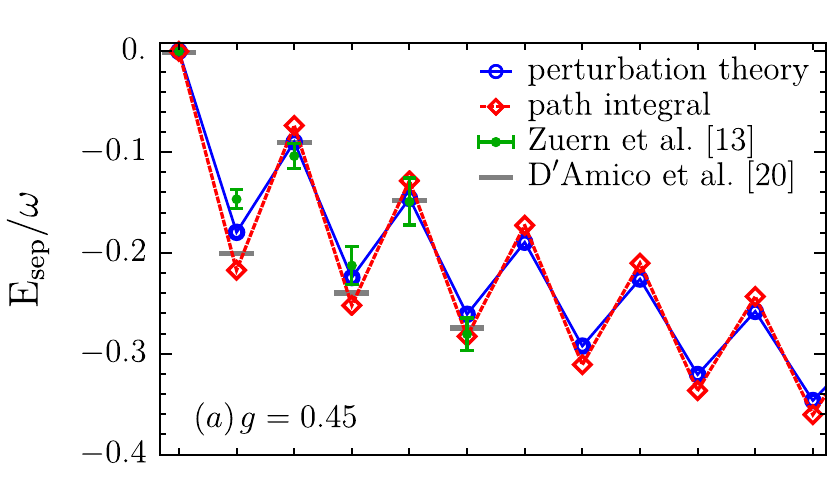}\label{fig:Esep}\vspace{-0.4cm}
\includegraphics{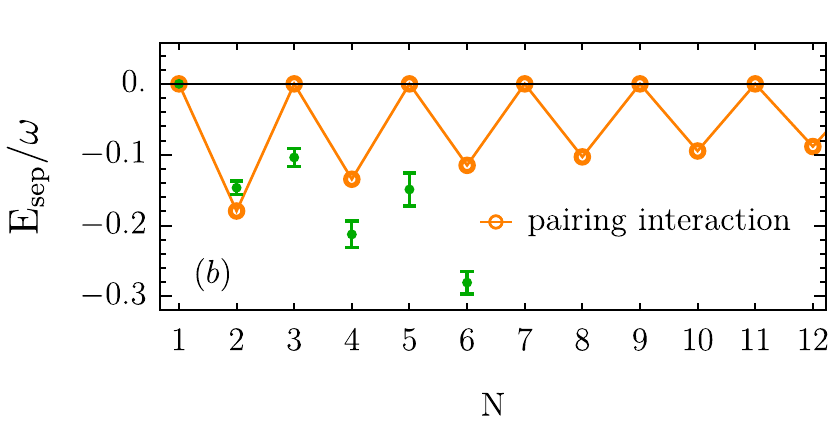}
\caption{(color online)
(a) Separation energy as a function of particle number $N$ for interaction strength $g_1/l_{\rm ho}\omega=0.45$. The leading-order perturbative solution of Eqs.~\eqref{eq:Esep1} and~\eqref{eq:Esep2} are indicated by blue circles and the path integral fluctuation correction [Eq.~\eqref{eq:fluccorr}] by red diamonds. Lines are a guide for the eye. Our result is in excellent agreement with the experimental results by Z\"urn {\it et al.}~\cite{zuern13}, which are shown as green error bars. In addition, we compare with results from exact diagonalization~\cite{damico15}, indicated by gray bars. (b) Orange line: separation energy including only pairing interactions. The Hartree density interaction is essential to fit the experimental data.
}
\label{fig:Esep}
\end{figure}

We proceed by analyzing the theory of Eq.~\eqref{eq:Hosc} in the weak-interaction limit to first order in the coupling $g$, applicable to the mesoscopic regime $\tilde{\Delta} \ll \omega$. To this leading order, the ground-state energy is given by the expectation value of Eq.~\eqref{eq:Hosc} with respect to the noninteracting ground state. The separation energy in Eq.~\eqref{eq:Esep} is thus
\begin{eqnarray}
E_{\rm sep}(2l+1) &=& -
g_1 l_{\rm ho}^{-1} \sum_{j=0}^{l-1} w_{jl} \label{eq:Esep1} \\
E_{\rm sep}(2l) &=& -
g_1 l_{\rm ho}^{-1} \sum_{j=0}^{l-1} w_{j(l-1)} \label{eq:Esep2} ,
\end{eqnarray}
which corresponds to the interaction energy of a single fermion in the outermost level  interacting with the fermions in the lower shells. In Eqs.~\eqref{eq:Esep1} and~\eqref{eq:Esep2}, we define the diagonal coupling $w_{ij} = w_{ijji}$, which can be determined in closed analytical form~\cite{kudla15}. Note that the perturbative interaction correction is due to a mean-field shift of the single-particle energies. In Fig.~\ref{fig:Esep}, we show the results of Eqs.~\eqref{eq:Esep1} and~\eqref{eq:Esep2} for the separation energy along with the experimental measurement~\cite{zuern13} (black error bars) for an interaction strength $g_1/\omega l_{\rm ho}=0.45$, which corresponds to the value used in the experiment~\cite{zuern13}. Remarkably, the perturbative result provides a very accurate description of the experimental data and is also in very good agreement with results from a numerical exact diagonalization of the Hamiltonian~\eqref{eq:Hosc} (gray bars)~\cite{damico15}. The parity parameter is
\begin{equation}
\Delta_P (\tilde{\Delta} \ll \omega) = \frac{g_1 w_{ll}}{2 \omega l_{\rm ho}} \omega \sim \tilde{\Delta} \sqrt{\frac{\omega}{\tilde{\Delta}}} , \label{eq:DeltaP}
\end{equation}
where $\tilde{\Delta}$ was defined in Eq.~\eqref{eq:macroparity}.

To gain insight into the physical mechanisms contributing to the separation energy and the parity parameter, we assume that for weak interactions, pairing takes place predominantly within a harmonic oscillator shell, i.e., that the ground-state properties can be described as excitations of paired levels: levels are either occupied by a pair of fermions or empty. This implies that only the interaction terms that connect two levels are retained: $w_{ij} = w_{iijj} = w_{ijij} = w_{ijji}$~\cite{*[{}] [{, Ch. 5.4.}] leggett06}. The effective Hamiltonian takes the form:
\begin{eqnarray}
&H_{\rm eff}& = \sum_{n\sigma} \varepsilon_n c_{n\sigma}^\dagger c_{n\sigma}^{} - 
g_1 l_{\rm ho}^{-1}\sum_{ij} w_{ij} c_{i\uparrow}^\dagger c_{i\downarrow}^\dagger c_{j\downarrow}^{} c_{j\uparrow}^{} \nonumber \\
&\quad& -
g_1 l_{\rm ho}^{-1}\sum_{i\neq j} w_{ij} c_{i\uparrow}^\dagger c_{j\downarrow}^\dagger c_{j\downarrow}^{} c_{i\uparrow}^{} . \label{eq:hameff}
\end{eqnarray}
There are two interaction terms: The first, which we call the {\it pairing term}, destroys a pair of spin-up and spin-down fermions in one oscillator level and creates a pair in a different one. The second, which we refer to as the {\it Hartree term}, does not create excitations but provides a density-dependent energy shift to the single-particle levels (note that the perturbative result is due to this type of interaction). A third possible interaction term which exchanges the spin between two simply occupied levels ({\it Fock term}) does not contribute to the balanced system that we consider. Compared to pairing models used for superconducting nanograins, the pairing interaction takes a more complicated level-dependent form and involves an additional Hartree term, which is in fact essential to describing the experimental data of Ref.~\cite{zuern13}. Figure~\ref{fig:Esep}(b) shows the leading order prediction for the separation energy only taking into account the pairing interaction. As is apparent from the figure, this prediction is in complete disagreement with the experiment. Note that the while the Hartree term is crucial for a correct description of the separation energy, it does not affect the parity parameter of Eq.~\eqref{eq:DeltaP} to leading order.

We obtain the ground-state energy for fixed particle number from the limit~\cite{matveev97}
\begin{equation}
\lim_{\beta \to \infty} \Omega_{\rm eff} = \min_N (E_N - \mu N) , \label{eq:gc}
\end{equation}
where $\Omega_\text{eff}$ is the free energy of the system obtained from 
the grand canonical partition function
${\cal Z} = e^{- \beta \Omega_{\rm eff}} = {\rm Tr} \, e^{- \beta (H_{\rm eff} - \mu N)}$ (where ${\rm Tr}$ denotes the trace over all many-particle eigenstates), i.e., the grand canonical ensemble projects onto a sector with definite particle number. However, because of the parity effect of Eq.~\eqref{eq:parityparameter}, the prescription~\eqref{eq:gc} only allows us to study configurations with even particle number. Nevertheless, we can relate the ground-state energy of a system with an odd number $2l+1$ of fermions to the ground-state energy of a system with an even number $2l$: since the Hamiltonian~\eqref{eq:hameff} only couples fully occupied or empty levels, the unpaired orbital of an odd-particle number state does not participate in the interaction and decouples; i.e., it effectively blocks a level from the Hilbert space. Hence~\cite{matveev97,vondelft01a,vondelft01b}
\begin{equation}
E_{2l+1} = \varepsilon_{l+1} + E_{2l}' ,
\end{equation}
where the energy $E_{2l}'$ is computed without the blocked level $l$. To analyze the effective theory of Eq.~\eqref{eq:hameff}, we eliminate the quartic interaction terms using a double Hubbard-Stratonovich transformation in both the density and pairing channel, which introduces three auxiliary fields $\Delta_i$, $\Delta_i^*$, and $K_i$. To this end, we define the operators $q_i^0 = \frac{1}{2} (c_{i\uparrow}^\dagger c_{i\uparrow}^{} + c_{i\downarrow}^\dagger c_{i\downarrow}^{})$, $q_i^+ = \frac{1}{2} (c_{i\uparrow}^\dagger c_{i\downarrow}^\dagger + c_{i\downarrow}^{} c_{i\uparrow}^{})$, and $q_i^- = \frac{i}{2} (c_{i\uparrow}^\dagger c_{i\downarrow}^\dagger - c_{i\downarrow}^{} c_{i\uparrow}^{})$. The Hamiltonian~\eqref{eq:hameff} takes the form $H - \mu N = T - V_+ - V_- - V_0$, where
\begin{eqnarray}
T &=& \displaystyle\sum_{j} \xi_j^0 n_{j}^{} + \frac{g_1 l_{\rm ho}^{-1}}{2} \sum_i w_{ii} , \label{eq:defT} \\ 
V_\alpha &=& \displaystyle g_1l_{\rm ho}^{-1} \sum_{ij} w_{ij} q_i^\alpha q_j^\alpha , \label{eq:defV}
\end{eqnarray}
with $\xi_j^0 = \varepsilon_j - \mu$. The last constant term in Eq.~\eqref{eq:defT} arises from a fermion commutator. Since the symmetric matrix $(w^{-1})_{ij}/g_1 l_{\rm ho}^{-1}$ is positive definite, the four-fermion interaction terms can be reduced using three standard Hubbard-Stratonovich transformations~\cite{negele98} for the operators $q^\alpha_i$ introducing conjugate real fields $x^\alpha_i$, where $\alpha = 0,\pm$.
Identifying $K_j = x_j^0$  and $\Delta_j = x_j^+ + i x_j^-$, the partition function reads:
\begin{align}
&{\cal Z}= \int \biggl[\frac{1}{ \cal N} \prod_{\tau,i} {\cal D}\Delta_i(\tau) {\cal D}\Delta_i^*(\tau) {\cal D}K_i(\tau) \biggr]\, {\rm Tr} \bigl[ U_\Delta(\beta,0) \bigr]  \nonumber \\
&
\displaystyle\times \exp\bigg[
- \beta C
- \int_0^\beta d\tau \, \sum_{ij} \frac{(w^{-1})_{ij}}{g_1 l_{\rm ho}^{-1}} \left(\Delta_i^* \Delta_j + K_i K_j\right)
 \bigg]
, \label{eq:ac3}
\end{align}
where $C = \sum_i (\frac{g_1
}{2 l_{\rm ho}} w_{ii} + \xi_i)$ with $\xi_j = \xi_j^0 - K_j$, $\mathcal{N}$ is the path integral normalization
\begin{eqnarray}
&&{\cal N} = \int \biggl[\prod_{\tau,i} {\cal D}\Delta_i(\tau) {\cal D}\Delta_i^*(\tau) {\cal D}K_i(\tau) \biggr] \nonumber \\
&& \times \exp\biggl[- \int_0^\beta d\tau \, \sum_{ij} \frac{(w^{-1})_{ij}}{g l_{\rm ho}^{-1}} \left(\Delta_i^* \Delta_j + K_i K_j\right) \biggl] ,
\end{eqnarray}
and
\begin{equation}
U_\Delta(\beta,0) = T_\tau \exp\biggl[- \int_0^\beta d\tau \sum_{j} \chi_j^\dagger
\begin{pmatrix}
\xi_j & -\Delta_j \\
- \Delta_j^* & - \xi_j
\end{pmatrix}
\chi_j^{}\biggr]
\end{equation}
with $\chi_j = (c_{j\uparrow}^{}, c_{j\downarrow}^\dagger)^T$~\cite{Galitski10}.

We first consider the saddle-point approximation and minimize the Euclidean action in Eq. (\ref{eq:ac3}) with respect to $K_i$ and $\Delta_i$. To this end, we first integrate out the fermions in the partition function, which gives the effective action
\begin{eqnarray}
&&S_{\rm eff}[\{\Delta_j, K_j\}] = - \sum_j {\rm tr} \ln [- G_{0,j}^{-1}] \nonumber \\
&&+ \int_0^\beta d\tau \, \biggl\{ \sum_{ij} \frac{(w^{-1})_{ij}}{g_1 l_{\rm ho}^{-1}} K_i K_j + \sum_i \xi_i + \frac{g_1}{2} \sum_i w_{ii} \biggr\} , \label{eq:effacsaddle}
\end{eqnarray}
where the trace runs over Matsubara indices. The matrix element of $G_{0,j}$ is given by $\langle i\omega_n | G_{0,j} | i\omega_{n'} \rangle = \delta_{n,n'} G_{0,j}(i\omega_n) = \delta_{n,n'} \bigl[i \omega_n - \xi_j \sigma_3
\bigr]^{-1} = \delta_{n,n'} \frac{i \omega_n + \xi_j \sigma_3}{(i\omega_n)^2 - \xi_j^2}$. Varying the action $S_{\rm eff}$ in Eq.~\eqref{eq:effacsaddle} with respect to $K_j$, $\Delta_j$, and $\mu$, we obtain the mean-field saddle-point solution defined by $K_i=\frac{g_1}{2 l_{\rm ho}} \sum_j w_{ij} \left(1 - \frac{\xi_j}{E_j}\right)$ and $\Delta_i = \frac{g_1 \omega}{l_{\rm ho}} \sum_j w_{ij} \frac{\Delta_j}{2 E_j}$, where $E_j =\sqrt{\xi_j^2 + \Delta_j^2}$. The solution of these equations determines the value of the Hartree field $K_i$ and the gap $\Delta_i$ at the saddle point. Note that for an even particle number, the saddle-point equations correspond to the solution of a BCS pairing ansatz~\cite{karpiuk04,swislocki08,kudla15} with coherence factors $v_i^2=1-u_i^2 = [1 -(\xi_i-K_i)/E_i]/2$.

For small $\gamma$, the only saddle-point solution for $\Delta_i$ corresponds to $\Delta_i=0$, which implies vanishing off-diagonal long-range order and absence of superfluidity in the weak-coupling regime. This is the fluctuation-dominated regime where $\tilde{\Delta}$ defined in Eq.~\eqref{eq:macroparity} is much smaller than the harmonic oscillator level spacing $\omega$~\cite{vondelft01a}. The Hartree fields are given by $K_i = g_1 l_{\rm ho}^{-1} \sum_{j=0}^{l-1} w_{ij}$, which correspond to the single particle energy shift computed to leading order in perturbation theory in $g$ using the noninteracting ground state of the Fermi gas. Using the identity in Eq.~(11), the saddle-point contribution to the ground-state energy of an system with even $N=2l$ particle number is $E_{2l} = 2 \sum_{j=0}^{l-1} \varepsilon_j - g_1 \sum_{i,j=0}^{l-1} w_{ij} + \frac{g_1}{2} \sum_i w_{ii}$. Interestingly, this is not equal to the perturbative ground-state energy. The last term arises from the commutator term in Eq.~\eqref{eq:defT}.  Despite the saddle point $\Delta_i$ being zero, fluctuations of the pairing field can make an important contribution, and they are computed in the remainder of this paper. It turns out that these pairing fluctuations contain a ${\cal O}(g_1)$ correction that cancels the last term in the saddle-point contribution to $E_{2l}$ and reproduces the perturbative result.

To consider the effect of fluctuations around the saddle-point solution. We write $K_i \to K_i + \delta K_i$ and $\Delta_i \to \delta \Delta_i$, and expand the action in Eq.~\eqref{eq:ac3} to second order in $\delta K_i$ and $\delta \Delta_i$. It turns out that there is no correction due to fluctuations of the Hartree fields $\delta K_i$. The partition function can be written as ${\cal Z} = {\cal Z}_{\rm sp} {\cal Z}_{\delta \Delta}$, with ${\cal Z}_{\rm sp}$ the saddle-point contribution and ${\cal Z}_{\delta \Delta} = [\det \Gamma]^{-1}$ the resulting quadratic functional integral in $\delta \Delta_i$, which can be exactly evaluated in terms of the functional determinant~\cite{matveev97,negele98}. To derive this result, we expand the perturbation in Matsubara space $\delta K_i(\tau) = \frac{1}{\sqrt{\beta}} \sum_{i\omega_n} e^{- i \omega_n \tau} \delta K_i(i\omega_n)$, and $\delta \Delta_i(\tau) = \frac{1}{\sqrt{\beta}} \sum_{i\omega_n} e^{- i \omega_n \tau} \delta \Delta_i(i\omega_n)$. The effective action is
\begin{eqnarray}
&&S_{\rm eff}[\{\theta_j\}] = - \sum_j {\rm tr} \ln [- G_{0,j}^{-1}] - \sum_j {\rm tr} \ln [1 - G_{0,j}^{} V_j^{}] \nonumber \\
&&+ \int_0^\beta d\tau \, \biggl\{\sum_{ij} \frac{(w^{-1})_{ij}}{g_1 l_{\rm ho}^{-1}} \Bigl[\delta \Delta_i^* \delta \Delta_j + (K_i+\delta K_i) \nonumber \\
&&\qquad \times (K_j+\delta K_j)\Bigr] + \sum_j (\xi_j - \delta K_j)\biggr\} , \label{eq:Sml}
\end{eqnarray}
where we separate a fluctuation part $V_j$ from the Green?s function $G_{0,j}$. The matrix element of $V_j$ is given by
\begin{eqnarray}
&&\langle i\omega_n | V_j | i\omega_{n'} \rangle = V_j(i\omega_{n} - i\omega_{n'}) \nonumber \\
&&= - \frac{1}{\sqrt{\beta}} \Bigl[\delta K_j(i\omega_{n} - i\omega_{n'}) \sigma_3 + \delta \Delta_j(i\omega_{n} - i\omega_{n'}) \sigma^+ \nonumber \\
&&\qquad + \delta \Delta_j^*(i\omega_{n} - i\omega_{n'}) \sigma^-\Bigr] .
\end{eqnarray}
Using
\begin{equation}
- \sum_j {\rm tr} \ln [1 - G_{0,j}^{} V_j^{}] = \sum_j \sum_{n=0}^\infty \frac{1}{j} {\rm tr} [G_{0,j}^{} V_j^{}]^n , \label{eq:explog}
\end{equation}
we expand the effective action~\eqref{eq:Sml} to second order in $V_j$. The functional integral in terms of $\delta K_j$ and $\delta \Delta_j$ can then be performed exactly. The partition function involving $\delta K_j$ reads:
\begin{widetext}
\begin{eqnarray}
&&{\cal Z}_{\delta K} = \frac{1}{{\cal N}_K} \int \biggl[\prod_k {\cal D} \delta K_k\biggr] \nonumber \\
&&\quad \qquad \times \exp\bigg[- \sum_{i\omega_n} \sum_{ij} \frac{(w^{-1})_{ij}}{g_1 l_{\rm ho}^{-1}}\delta K_i(- i\omega_n) \delta K_j(i\omega_n) - \sqrt{\beta} \sum_{i} \delta K_i(i\omega_m=0) \biggl(2 \sum_j \frac{(w^{-1})_{ij}}{g_1 l_{\rm ho}^{-1}} K_j + \frac{\xi_j}{|\xi_j|} - 1\biggr)\biggr] .
\end{eqnarray}
However, the zero-frequency contribution (second term) vanishes since $K_j$ is given by $K_j = g_1 l_{\rm ho}^{-1} \sum_{i=0}^{l-1} w_{ij}$. The remaining quadratic term is irrelevant since it does not involve any single-particle energies. The partition function involving $\delta \Delta_j$ reads (discarding an irrelevant constant term)~\cite{negele98,matveev97}:
\begin{eqnarray}
{\cal Z}_{\delta \Delta} &&= \frac{1}{{\cal N}_\Delta} \int \biggl[\prod_k {\cal D} \delta \Delta_k^* {\cal D} \delta \Delta_k\biggr] \exp\bigg[- \sum_{i\omega_n} \sum_{ij} \delta \Delta_i^*(- i\omega_n) \left(\frac{(w^{-1})_{ij}}{g_1 l_{\rm ho}^{-1}} + \delta_{ij} \frac{{\rm sgn} \, \xi_j}{i\omega_{n} - 2 \xi_j} \, \right) \delta \Delta_j(i\omega_n)\biggr] \nonumber \\
&&= \prod_{i\omega_n} {\rm det}^{-1}\biggl[\delta_{ij} + g_1 l_{\rm ho}^{-1} w_{ij} \frac{{\rm sgn} \, \xi_j}{i\omega_{n} - 2 \xi_j}\biggr] = \prod_j \frac{\sinh \beta \xi_j}{\sinh \beta \tilde{\xi}_j} ,
\end{eqnarray}
\end{widetext}
where by $\{2 \tilde{\xi}_j\}$ we denote the eigenvalues of $A_{ij} = 2 \xi_j \delta_{ij} - g_1 w_{ij} \, {\rm sgn} \, \xi_j $, i.e., $\det(2 \tilde{\xi}_k I - A) = 0$, or, respectively,
\begin{eqnarray}
&&\det\biggl(\delta_{ij} + \frac{g_1 l_{\rm ho}^{-1}}{2} \frac{w_{ij} \, {\rm sgn} \, \xi_j}{\tilde{\xi}_k - \xi_j}\biggr) \nonumber \\
&&\quad= 1 + \frac{g_1 l_{\rm ho}^{-1}}{2} \sum_i \frac{w_{ii} \, {\rm sgn} \, \xi_i}{\tilde{\xi}_k - \xi_i} = 0 ,
\end{eqnarray}
where the second term holds for small corrections. Writing $\tilde{\xi}_j = \xi_j + \delta \xi_j$ and expanding in $\delta \xi_j$, the fluctuation correction to the free energy at zero temperature is:
\begin{equation}
\delta \Omega = - \frac{1}{\beta} \ln {\cal Z}_{\delta\Delta} = \sum_j \delta \xi_j \, {\rm sgn} \, \xi_j , \label{eq:fluccorr}
\end{equation}
where
\begin{equation}
\delta \xi_i = - \frac{g_1 w_{ii} \, {\rm sgn} \, \xi_i}{2 l_{\rm ho}}  \biggl(1 - \dfrac{g_1}{2l_{\rm ho}} \displaystyle \sum_{j\neq i} \dfrac{w_{jj} \, {\rm sgn} \, \xi_j}{\xi_j - \xi_i}\biggr)^{-1} . \label{eq:flucorr}
\end{equation}
Hence, $\delta E_{2l} = \sum_j \delta \xi_j \, {\rm sgn} \xi_j$ and $\delta E_{2l+1} = \sum_{j\neq l} \delta \xi'_j \, {\rm sgn} \xi'_j$, where $\xi'$ and $\delta \xi'$ are computed as in Eq.~\eqref{eq:flucorr} with the $l$-th level excluded. 

The fluctuation correction to the separation energy and the parity parameter can be read off directly from the definitions~\eqref{eq:parityparameter} and~\eqref{eq:Esep}. Note that the combined saddle-point and fluctuation correction contains the leading order perturbative result (see Fig. ~\ref{fig:Esep}, where the separation energy is indicated by the red dashed line and the diamond symbol). There is a small quantitative correction which improves the agreement with the exact diagonalization results by D'Amico {\it et al.}~\cite{damico15}. The fluctuation correction~\eqref{eq:flucorr} is similar to the one encountered in superconducting nanograins in the limit where the superconducting gap $\Delta$ is much smaller than the level spacing $\delta \epsilon$~\cite{matveev97}.

Interestingly, our analytical procedure also allows us to identify the critical particle number that marks the crossover between the few-body regime $\tilde{\Delta} \ll \omega$ and the many-body regime $\tilde{\Delta} \gg \omega$. The boundary of the mesoscopic regime is marked by a breakdown of the expansion~\eqref{eq:flucorr}. For large particle number, we can replace the harmonic oscillator matrix element by its semiclassical expression $w_{jj} \sim 1/\sqrt{j}$ and convert the summation to an integral. This gives
\begin{equation}
\frac{g_1}{2 l_{\rm ho}} \sum_{j\neq i} \frac{w_{jj} \, {\rm sgn} \, \xi_j}{\xi_j - \xi_i} \sim \frac{g_1}{\omega l_{\rm ho}} \sqrt{\frac{4}{N}} \ln 2 N \sim \gamma \ln 2 N ,
\end{equation}
indicating that by successively increasing particle number, the few-body expansion loses validity at a critical particle number $N_c \sim e^{1/\gamma}$. In this case, the bulk parity parameter~\eqref{eq:macroparity} is comparable to the level spacing $\omega$, which is a corresponding criterion as in superconducting nanograins~\cite{vondelft01a}. Note that while the few-body to many-body crossover is manifested in the parity parameter at leading order, the ground-state energy is dominated by a Hartree mean-field term and will be less sensitive to the crossover. From the perspective of superconducting nanograin systems, such a predominance of the Hartree contribution over the fluctuation correction is an unexpected effect \cite{matveev97,vondelft01a,vondelft01b}. Therefore, our findings prompt a revision of both the theoretical modeling of nanograins and the related experimental results \cite{Lagarge93,ralph95,black96,ralph97}. 

While the BCS pairing model can be solved exactly~\cite{richardson63,richardson64,richardson65,richardson66,sierra00,vondelft00,dukelsky00,dukelsky04}, this is not the case for the model~\eqref{eq:Hosc} or the reduced pairing Hamiltonian~\eqref{eq:hameff}. However, it would be interesting to explore if the theory could be approximated by a generalized Richardson-Gaudin model~\cite{dukelsky01}.
 
In summary, we have computed the ground-state energy, the separation energy, and the parity parameter for a trapped one-dimensional Fermi gas with weak attractive interaction. We have used an insightful path-integral formalism which allows us to make useful connections with other physical systems (i.e., mesoscopic superconductors). The parity parameter serves as an order parameter that displays a fundamentally distinct behavior in the mesoscopic and macroscopic regimes, and we establish that the mesoscopic description persists for a wide range of particle number. Our results provide a quantitative description of the recent experiment~\cite{zuern13}. A path integral treatment indicates that the ground-state energy and the parity effect are dominated by a Hartree mean field contribution, with BCS pairing fluctuations providing a subleading correction to this result.

\begin{acknowledgments}
We thank M. Rontani for sharing the data of Ref.~\cite{damico15}. This work is supported by LPS-MPO-CMTC, JQI-NSF-PFC, and ARO-MURI (J.H.), CONICET-PIP 00389CO (A.M.L). V.G. acknowledges support from DOE-BES DESC0001911, Australian Research Council, and Simons Foundation.
\end{acknowledgments}

\bibliography{bib}

\end{document}